\documentclass[iop]{emulateapj}

\usepackage{graphicx}
\usepackage{amsmath}
\usepackage{natbib}
\usepackage{mathptmx}
\usepackage{apjfonts}
\usepackage{times}
\usepackage[colorlinks=true,urlcolor=blue,citecolor=blue,linkcolor=blue]{hyperref}
\usepackage{aas_macros}
\usepackage{tikz}

\bibpunct{(}{)}{;}{a}{}{,}

\usepackage[position=t,singlelinecheck=on,font={rm,bf,up},font=large]{subfig}
\RequirePackage{color}
\newcommand{\sunrise}{\textsc{Sunrise}}
\newcommand{\sufi}{\textsc{SuFI}}
\definecolor{darkgreen}{rgb}{0.0, 0.3, 0.0}
\definecolor{al}{rgb}{0.6, 0.3, 0.0}

\newcommand{\kms}{km/s}
\newcommand{\Mypm}{\mathbin{\tikz [x=1.4ex,y=1.4ex,line width=.1ex] \draw (0.0,0) -- (1.0,0) (0.5,0.08) -- (0.5,0.92) (0.0,0.5) -- (1.0,0.5);}}%

\newcommand{\caiih}{Ca\,\textsc{ii}\,H}
\newcommand{\caiik}{Ca\,\textsc{ii}\,K}
\newcommand{\fig}[1]{Fig.~\ref{#1}} 
\newcommand{\sref}[1]{Sect.~\ref{#1}} 

\shorttitle{Oscillations on width and intensity of slender \caiih{} fibrils}
\shortauthors{Gafeira et al.}

\begin{document}

\title{Oscillations on width and intensity of slender \caiih{} fibrils from \sunrise/\sufi{}}

\author{R.~Gafeira\hyperlink{}{\altaffilmark{1}}}
\author{S.~Jafarzadeh\hyperlink{}{\altaffilmark{2}}}
\author{S.~K.~Solanki\hyperlink{}{\altaffilmark{1,3}}}
\author{A.~Lagg\hyperlink{}{\altaffilmark{1}}}
\author{M.~van~Noort\hyperlink{}{\altaffilmark{1}}}
\author{P.~Barthol\hyperlink{}{\altaffilmark{1}}}
\author{J.~Blanco~Rodr\'{i}guez\hyperlink{}{\altaffilmark{4}}}
\author{ J.~C.~del~Toro~Iniesta\hyperlink{}{\altaffilmark{5}}}
\author{A.~Gandorfer\hyperlink{}{\altaffilmark{1}}}
\author{L.~Gizon\hyperlink{}{\altaffilmark{1,6}}}
\author{J.~Hirzberger\hyperlink{}{\altaffilmark{1}}}
\author{M.~Kn\"{o}lker\hyperlink{}{\altaffilmark{7}}}
\author{D.~Orozco~Su\'{a}rez\hyperlink{}{\altaffilmark{5}}}
\author{T.~L.~Riethm\"{u}ller\hyperlink{}{\altaffilmark{1}}}
\author{W.~Schmidt\hyperlink{}{\altaffilmark{8}}}

\affil{\altaffilmark{1}\hspace{0.2em}Max Planck Institute for Solar System Research, Justus-von-Liebig-Weg 3, 37077 G\"{o}ttingen, Germany; \href{mailto:gafeira@mps.mpg.de}{gafeira@mps.mpg.de}\\
\altaffilmark{2}\hspace{0.2em}Institute of Theoretical Astrophysics, University of Oslo, P.O. Box 1029 Blindern, N-0315 Oslo, Norway\\
\altaffilmark{3}\hspace{0.2em}School of Space Research, Kyung Hee University, Yongin, Gyeonggi 446-701, Republic of Korea\\
\altaffilmark{4}\hspace{0.2em}Image Processing Laboratory, University of Valencia, P.O. Box 22085, E-46980 Paterna, Valencia, Spain\\
\altaffilmark{5}\hspace{0.2em}Instituto de Astrof\'{i}sica de Andaluc\'{i}a (CSIC), Apartado de Correos 3004, E-18080 Granada, Spain\\
\altaffilmark{6}\hspace{0.2em}Institut f\"ur Astrophysik, Georg-August-Universit\"at G\"ottingen, Friedrich-Hund-Platz 1, 37077 G\"ottingen, Germany\\
\altaffilmark{7}\hspace{0.2em}High Altitude Observatory, National Center for Atmospheric Research, \footnote{The National Center for Atmospheric Research is sponsored by the National Science Foundation.} P.O. Box 3000, Boulder, CO 80307-3000, USA\\
\altaffilmark{8}\hspace{0.2em}Kiepenheuer-Institut f\"{u}r Sonnenphysik, Sch\"{o}neckstr. 6, D-79104 Freiburg, Germany
}

\begin{abstract}

We report the detection of oscillations in slender \caiih{} fibrils (SCFs) from high-resolution observations acquired with the \sunrise{} balloon-borne solar observatory. The SCFs show obvious oscillations in their intensity, but also their width. The oscillatory behaviors are investigated at several positions along the axes of the SCFs. A large majority of fibrils show signs of oscillations in intensity. Their periods and phase speeds are analyzed using a wavelet analysis. 
The width and intensity perturbations have overlapping distributions of the wave period.

The obtained distributions have median values of the period of $32\pm17$\,s and $36\pm25$\,s, respectively. We find that the fluctuations of both parameters propagate in the SCFs with speeds of ${11}^{+49}_{-11}$\,\kms{} and ${15}^{+34}_{-15}$\,\kms{}, respectively. Furthermore, the width and intensity oscillations have a strong tendency to be either in anti-phase, or, to a smaller extent, in phase. This suggests that the oscillations of both parameters are caused by the same wave mode and that the waves are likely propagating. Taking all the evidence together, the most likely wave mode to explain all measurements and criteria is the fast sausage mode.

\vspace{1mm}
\end{abstract}


\section{Introduction}
\label{sec:intro}

Magnetohydrodynamic (MHD) waves have been observed in various plasma structures in the solar atmosphere, particularly in elongated features in the solar chromosphere and in the corona \cite[for recent reviews see, e.g.,][]{Banerjee2007,Zaqarashvili2009,Mathioudakis2013,Jess2015b}. According to the theory of MHD oscillations, the waves may appear as a single mode, or as a combination of several modes (i.e., kink, sausage, torsional, or longitudinal) with distinct properties and different observational signatures \citep{Edwin1983}. These waves are often excited at photospheric heights (e.g. by granular buffeting, \citealt{Evans1990}, or vortex motion, \citealt{Kitiashvili2011}), and either propagate away from their source, or form a standing oscillation.

While transverse waves, such as kink or global Alfv\'enic modes cause swaying of a flux tube or of an elongated feature, sausage-mode oscillations result in a periodic axisymmetric expansion and contraction of the structure at one position. Torsional or twisting motions are associated with torsional  Alfv\'{e}n waves propagating along the axis of fibrillar structures \citep[e.g.,][]{Spruit1982,Solanki1993}.

The different wave modes in elongated structures have been mostly observed at coronal heights in, e.g., coronal loops \cite[e.g.,][see \citealt{lrsp-2005-3} for a review]{Aschwanden1999,DeMoortel2002,2002ApJ...574L.101W,2004A&A...421L..33W,Tomczyk2007,Srivastava2008,Nakariakov2012}, as well as at the upper chromospheric levels in features such as filaments, fibrils, mottles and spicules \citep{DePontieu2007,Lin2007,Pietarila2011,Okamoto2011,Tsiropoula2012}.

Images recorded in the \caiik{} passband (with a filter width of 1.5\,\AA) with the  1-m Swedish Solar Telescope (SST; \citealt{Scharmer2003}) in an active region close to the solar disk center revealed the presence of slender, bright fibrils, extending seemingly horizontally in the lower chromosphere \citep{Pietarila2009}. Only recently, the second flight of the 1-m balloon-borne solar observatory \sunrise{} \citep{Solanki2010,Barthol2011,Berkefeld2011,Solanki2016} provided us with a high quality, seeing-free time-series of \caiih{} images (with a filter width of 1.1\,\AA). The relatively long duration of the observations (one hour) in an active region (close to disk center) enabled a thorough study of properties of the slender fibrils \citep{Gafeira2016}. In addition, \citet{Jafarzadeh2016c} have found ubiquitous transverse waves in the slender \caiih{} fibrils (SCFs) in \sunrise{} data. The SCFs have been shown to map the magnetic fields in the low solar chromosphere \citep{Jafarzadeh2016b}. Earlier, \citep{Pietarila2009} had shown that the fibrils, or the magnetic canopy outlined by them, either suppressed oscillations or channeled low frequency oscillations into the chromosphere, depending on their location. 

In this paper we investigate width and intensity oscillations in the SCFs observed with \sunrise{}. Periods of the fluctuations in individual SCFs are determined and the phase speed of the waves propagating along the thin structures are quantified (\sref{sec:analysis}). We discuss our results in \sref{sec:conclusions}, where we also conclude that the observed oscillations are likely manifestations of sausage waves traveling along the fibrils.

\section{Observations}
\label{sec:obs}

For the present study we use the data set described in \citet{Gafeira2016}, \citet{Jafarzadeh2016c}, and \citet{Solanki2016}. The data set includes high spatial and temporal resolution observations of an active region obtained in the \caiih{} passband (with a full width at half maximum, FWHM, of $\approx1.1$\,\AA) of the \sunrise{} Filter Imager \cite[\sufi{},][]{Gandorfer2011} on the 1-meter \sunrise{} balloon-borne solar observatory \cite[]{Solanki2010,Barthol2011,Berkefeld2011} during its second science flight \cite[]{Solanki2016}.
The observations were obtained between 23:39~UT on 2013 June 12 and 00:38~UT on 2013 June 13 with a cadence of 7\,s. These observations covered a part of NOAA AR 11768, mainly its following polarity that was dominated by a series of pores, some of which lie at least partly within the \sufi{} field-of-view (FOV). The FOV also contained a number of plage elements and two granulation-scale flux emergences \citep[e.g.,][]{centeno2016}. The FOV was centered at $\mu=\cos\theta=0.93$, where $\theta$ is the heliocentric angle.

\begin{figure}[ht]
   \centering
	\includegraphics[width=9cm]{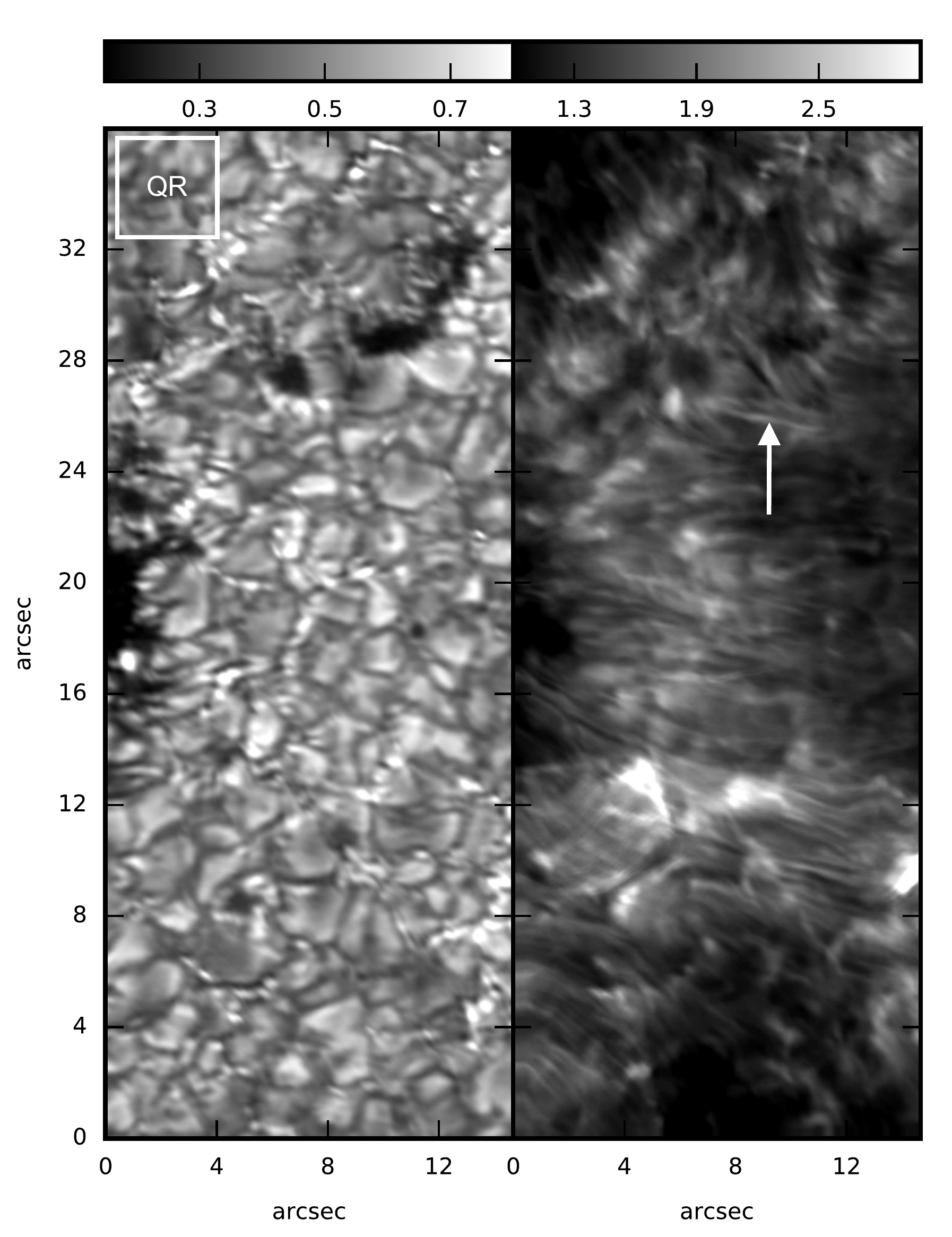}
   \caption{Right: Example of a \caiih{} image recorded by \sunrise{}/\sufi{}. Left: An image recorded at 300\,nm, aligned with the \caiih{} filtergram. The white arrow in the right panel indicates a sample slender \caiih{} fibril (SCF). The white box marks the quiet region (QR) used to normalize the intensities in the \caiih{} and 300\,nm images.}
   \label{Fig:obs}
\end{figure}

\fig{Fig:obs} illustrates a \caiih{} image (right panel) along with its co-spatial and co-temporal photospheric filtergram (left panel) recorded at 300\,nm with the \sunrise{}/\sufi{} instrument. The arrow on the \caiih{} image marks an example SCF studied here. All the intensity values where normalized to the mean intensity of the relatively quiet region, indicated by the white box in the top left corner of the \fig{Fig:obs}.

The \caiih{} image has both, a photospheric and a low chromospheric component. The latter is strongly enhanced in an active region \cite[]{Danilovic2014,Jafarzadeh2016b}. The fibrils dominating much of the \caiih{} image are expected to be located in the lower chromosphere, as none of the photospheric channels on \sunrise{} even remotely shows any signs of such fibrils \cite[see][]{Jafarzadeh2016b}.

\section{Analysis and results}
\label{sec:analysis}

\begin{figure}[ht]
   \centering
	\includegraphics[width=8cm]{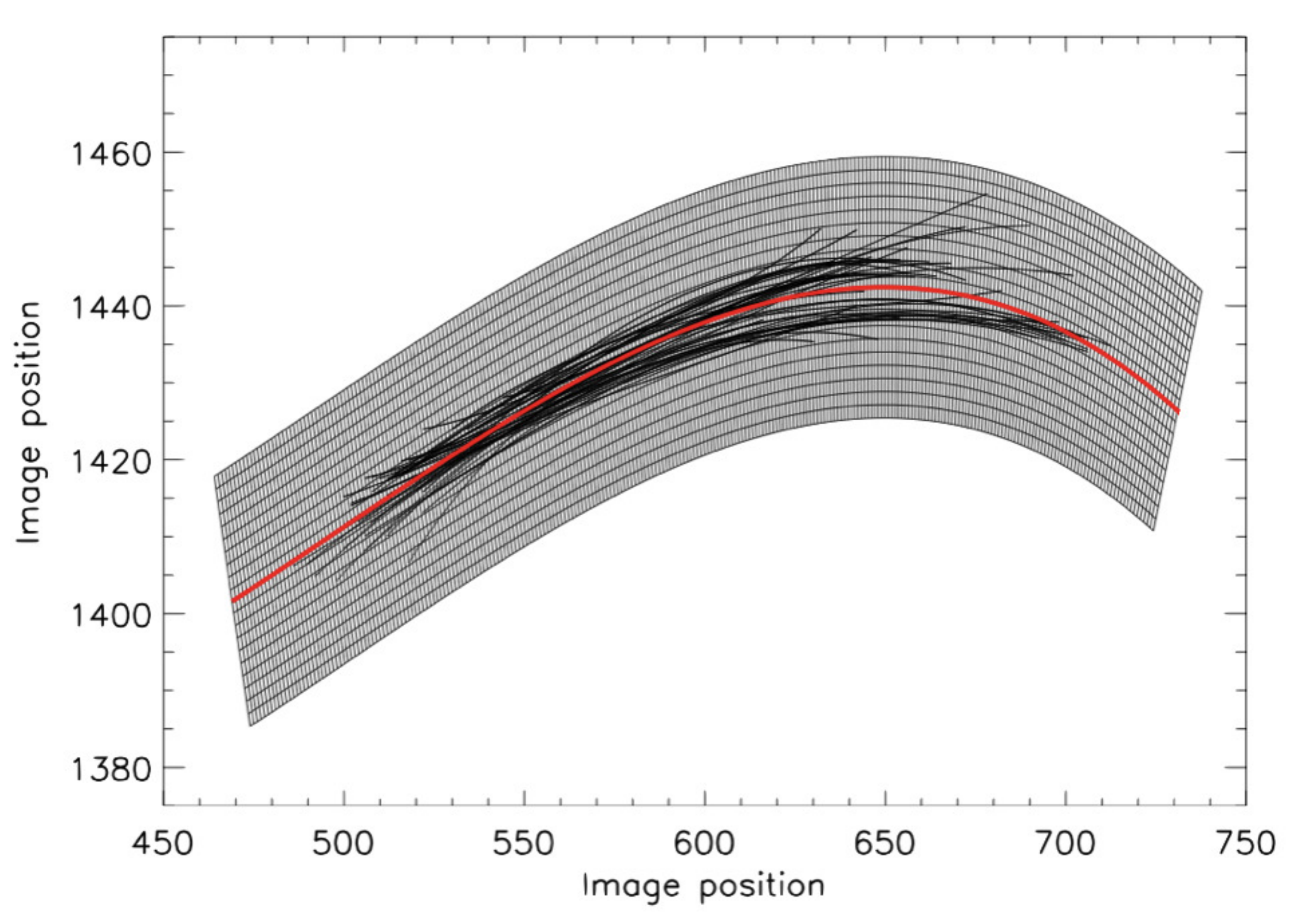}
   \caption{Illustration of backbones of a sample SCF (at different times and averaged) and the grid associated with the SCF. The individual backbones of this fibril determined in the images recorded at different times are represented by the individual black lines. The reference backbone is represented by the red line and the mesh is shown by the black grid.}
   \label{mesh_con}
\end{figure}

To analyze both, the intensity and width variations of the SCFs with time, we first have to extract the fibrils from the \caiih{} filtergrams. We follow the
identification and tracking method described in detail in \citet{Gafeira2016}.
In a first step, this method defines a binary mask of all the fibrils. This mask is obtained by applying an unsharp masking and an adaptive histogram equalization method to the intensity images to increase their contrast. In these images, a threshold of 50\% of the maximum intensity defines the binary mask isolating the fibrils from the background. All features smaller than the diffraction limit of the telescope are discarded. For the temporal evolution of a fibril we require at least 10 pixels of the fibril to be visible at the same position in at least 5 out of 6 subsequent frames, corresponding to a minimum lifetime of 35\,s. In all these frames, a fibril backbone is defined as the line equidistant to the fibril's border. A second order polynomial is fitted to all backbones of the individual fibrils. Fibrils of complex shape, that are poorly fitted, are excluded from the analysis.
We extend this fitted curve by 0.3\arcsec{} in both end points, the approximate width of a fibril, to compensate for the reduction of the fibril to a single-pixel structure.
The resulting line is what we call the reference backbone (see red line in \fig{mesh_con}). For more details, we refer to \citet{Gafeira2016}.
This reference backbone is the central line for the mesh, displayed in \fig{mesh_con}. All other points of the mesh are calculated from equally spaced lines perpendicular to this reference backbone. 
This mesh is based on the fibril backbone, i.e. the temporal average of all fibrils, and is therefore time-independent,
allowing the study of the temporal variations of its brightness and width.
We use a mesh with a fixed total width of 1.2\arcsec{} for all the fibrils, while the length is the same as the reference backbone of each individual SCF. This procedure results in a total number of 598 detected SCF over the full data set with lifetimes of 35\,s or longer.

\begin{figure}[ht]
   \centering
   \includegraphics[width=9cm]{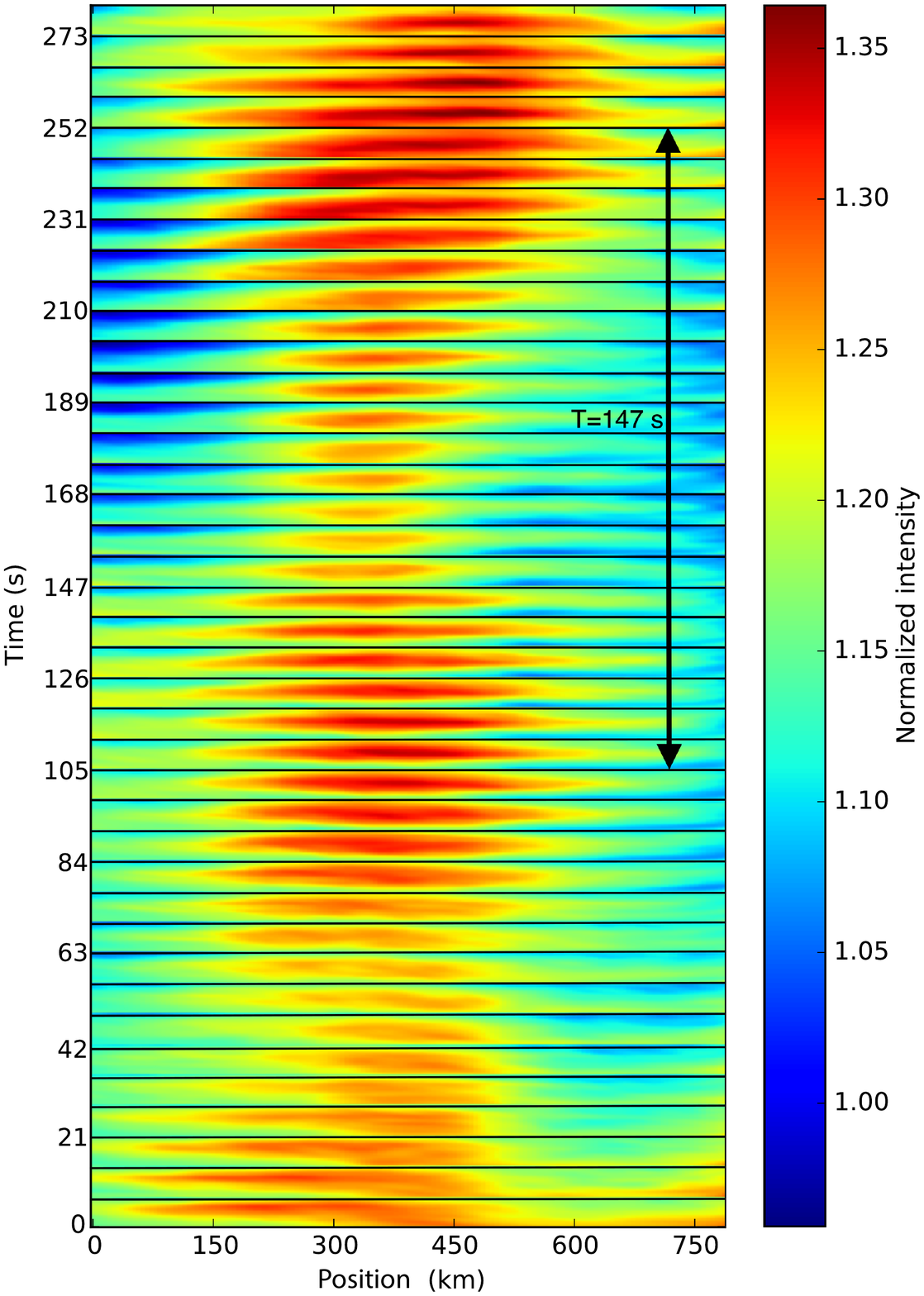}
   \caption{Temporal variation of a SCF. The images of a straightened fibril at different times are vertically stacked. Individual images, recorded every 7\,s, are separated by horizontal black lines. The vertical arrow indicates the period of the fibril's intensity fluctuation. The colour represents intensity, normalized to the mean value of the quiet region marked in \fig{Fig:obs}.}
   \label{dfmesh_ex1}
\end{figure}

Using this approach, we straightened each identified SCF in each image (observed at different times) by interpolating its intensity onto every point of the mesh. After these steps we can represent the SCF along a straight line, as shown in \fig{dfmesh_ex1} for an example fibril pictured at 40 time steps. Every black-framed box in \fig{dfmesh_ex1} includes the straightened SCF at a given time. This way of stacking the temporal snapshots of a fibril allows us to easily follow fluctuations of its intensity (and with some additional effort also of its width) at any location along the reference backbone of the SCF during its entire lifetime. Thus we can identify different types of oscillations/pulsations in these structures. The example shown in \fig{dfmesh_ex1} illustrates a clear fluctuation of the intensity with a period of approximately 147\,s (21 frames), indicated by the double-headed vertical arrow.

To inspect the fluctuations in both, the intensity relative to the mean intensity of the quiet Sun, and the width of the SCFs in detail, we evaluate the intensity at 17 positions along the backbone of the fibril (lying between 20 and 80 percent of the full length of the reference backbone measured from one of its ends), perpendicular to which we create artificial slits (that correspond to a given set of $x$ positions in the mesh frame). We then compute the position of the maximum intensity of the fibril, the intensity at this position, as well as the width of the fibril along each of these slits following the method outlined further below.

In some cases a second fibril may be present inside the mesh determined for one fibril, usually near the edge of the mesh. As a result, more than one local maximum is present along the artificial slit used for the determination of the fibril's width. In such cases, we choose the local maximum which is closer to the center of the mesh, i.e., the reference backbone of the fibril.

\begin{figure}[ht]
   \centering
   \includegraphics[width=9cm]{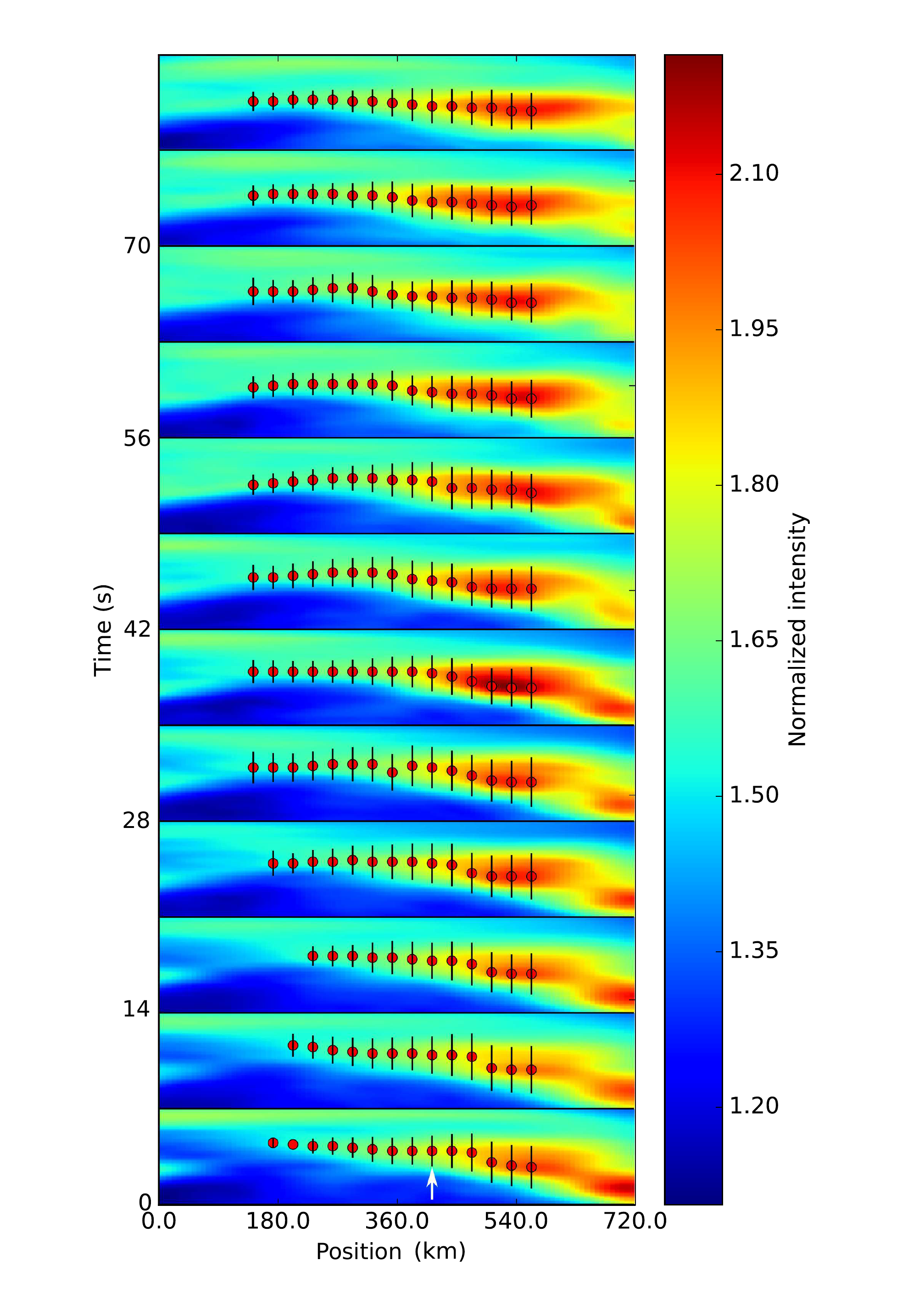}
   \caption{An example of intensity maxima and width detections along cuts perpendicular to the axis of a SCF. Plotted are vertically stacked images of a fibril recorded in \caiih{} observed at different times. Individual images, recorded every 7\,s, are separated by horizontal black lines. The red dots within a given image represent the locations of the fibril's maximum intensity along a series of cuts roughly perpendicular to the backbone of the fibril, while the vertical black lines indicate the width of the fibril at the same locations. The colour represents intensity, normalized to the mean value of the quiet region in the \sufi{} frame (marked in \fig{Fig:obs}). The white arrow in the lower part of the bottom image marks the location at which the oscillations plotted in \fig{int_wid} occurred. Note that this fibril is not the same as presented in \fig{dfmesh_ex1}.}
   \label{mesh_mp_w_ex1}
\end{figure}

The width of a fibril is computed by fitting a Gaussian function plus a linear background to the intensity profile perpendicular to the backbone. To minimize the influence of neighboring fibrils, the six points closest to the maximum intensity position are given 30\% higher weights. The FWHM of the fitted Gaussian defines the fibril width. In \fig{mesh_mp_w_ex1} we present an example of these measurements, where positions of the maximum intensity (red circles) and the width of the fibril (vertical black lines) are marked at various positions along the SCF (within each image) and at different times (from one image in the stack to the next). For better visibility, we have chosen a relatively short-lived SCF for clarity (i.e., we need to show fewer time-steps). The plot clearly shows that the width is bigger in the brighter part of the fibril. The way the width is determined (see above), is independent of the fibril's intensity, as long as the profile shape of the intensity perpendicular to the fibril's axis does not change and the fibril's intensity is higher than the background.
We note that the locations along the fibrils are determined only as long as the intensity along the SCF is larger than the average intensity of the image at each time-step. Therefore, the positions close to the left end of the example SCF shown in \fig{mesh_mp_w_ex1} are not detected at all times.

\subsection{Wavelet analysis}
\label{sec:wavelet}

We apply a wavelet analysis to characterize the temporal variation of the power spectrum of the width and intensity oscillations. We use the wavelet algorithm  described by \citet{Jafarzadeh2016a}. 
For the cases with a clear intensity and width oscillation, we also calculate cross power spectra, i.e., the multiplication of the wavelet power spectrum of the oscillation in a given quantity at one position along the fibril by the complex conjugate of the same wavelet power spectrum at a different location along the same fibril. 
This provides us with the phase differences between the consecutive positions in width and intensity oscillations, and hence, with the phase speed of the waves along the fibrils. Finally, we also determine the wavelet cross power spectrum between brightness and width oscillations that provides the phase difference between them.

In some cases, the determination of the maximum intensity and width of a fibril at some positions along the fibril at a given time is difficult, leading to gaps in some of the 17 positions along the fibril backbone. These gaps are filled by linearly interpolating in time to provide the wavelet analysis with equidistant data points. Such interpolations can result in overestimation of, e.g., periods of the oscillations. Gaps are sufficiently rare, however, that their influence turned out to be relatively insignificant. We find that 74\% of the fibrils display above 95\% confidence level (inside the cone of influence) oscillations in intensity, with on average 42\% of the cuts along each oscillating fibril displaying such an oscillation. Similarly, 82\% of the fibrils exhibit oscillations of the width, whereby 38\% of the cuts through the backbone of oscillating fibrils show the oscillations (on average).
For the fibrils displaying an oscillation with a sufficiently high confidence, the frequency at which the wavelet power spectrum has its strongest peak is taken as the period of the oscillation (within a given fibril). The most likely phase speed of the wave is determined in the same way from the wavelet-cross-power spectrum between different spatial locations along a fibril. Only the highest peaks that are above the 95\% confidence level and inside the cone of influence (i.e., frequency-time areas that are not influenced by the ends of the time-series) are considered.

\subsection{Statistics}
\label{sec:statistics}

\begin{figure}[ht]
   \centering
   \includegraphics[width=8cm]{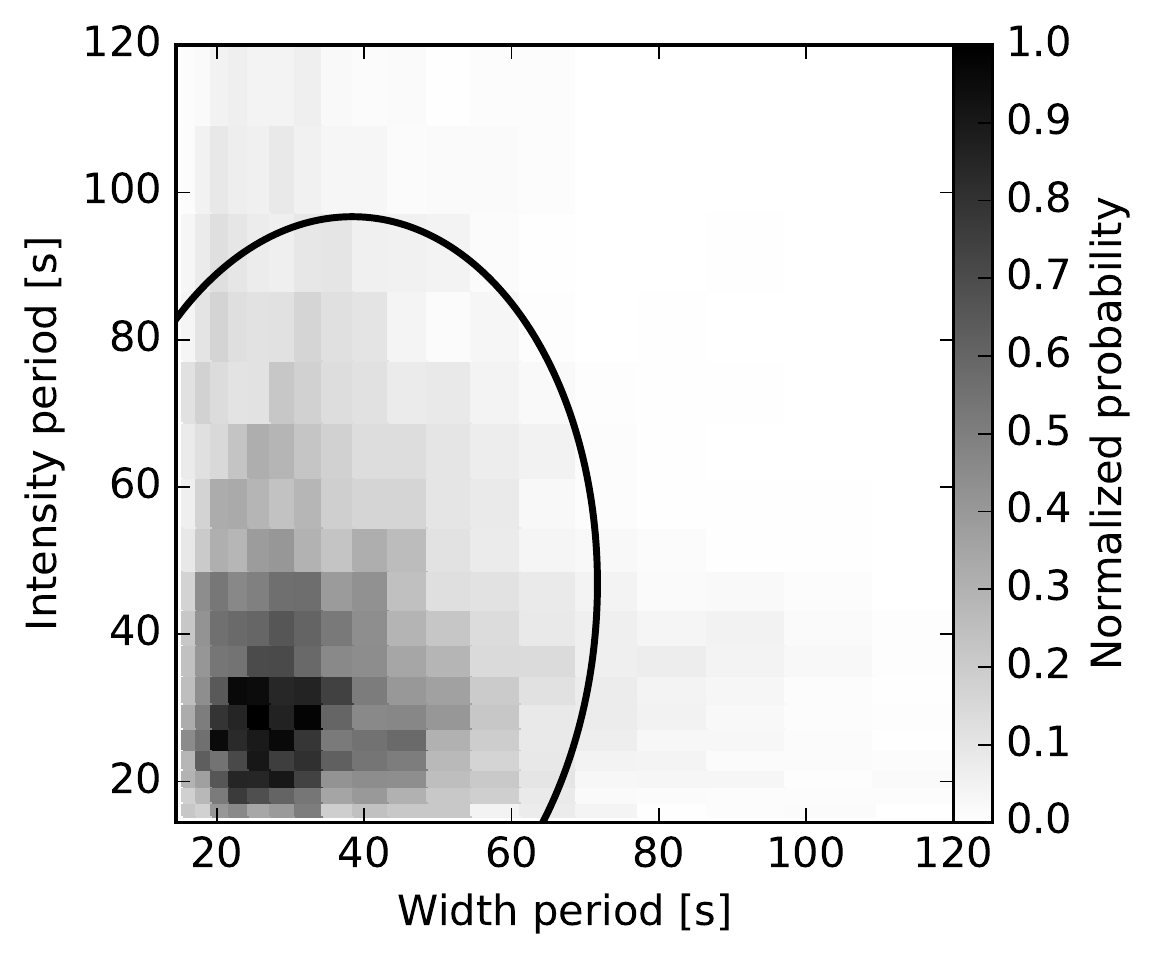}
   \caption{Two-dimensional histogram showing the relation between the periods of the width and the intensity oscillations in SCFs. The bin size follows the period resolution that is limited by the life time of the fibrils. The black curved line indicates the 95\% confidence level.
}
   \label{perhist}
\end{figure}

The two-dimensional histogram of intensity and width periods presented in \fig{perhist} demonstrates that most of the fibrils oscillate with periods between 20\,s and 40\,s in both quantities, with median values of $32\pm17$\,s and $36\pm25$\,s for the periods of the width and intensity oscillations, respectively. For a large fraction of the fibrils ($\approx$75\%) the periods in both quantities are similar.
For the phase speeds we obtain median values of ${11}^{+49}_{-11}$\,\kms{}  and ${15}^{+34}_{-15}$\,\kms{} for width and intensity oscillations, respectively, without any correlation between them.
These median periods are rather short, well below the cut-off frequency of the atmosphere, while the phase speeds are above the sound speed in the temperature minimum region and lower chromosphere.

\begin{figure}[ht]
   \centering
   \includegraphics[width=8.5cm]{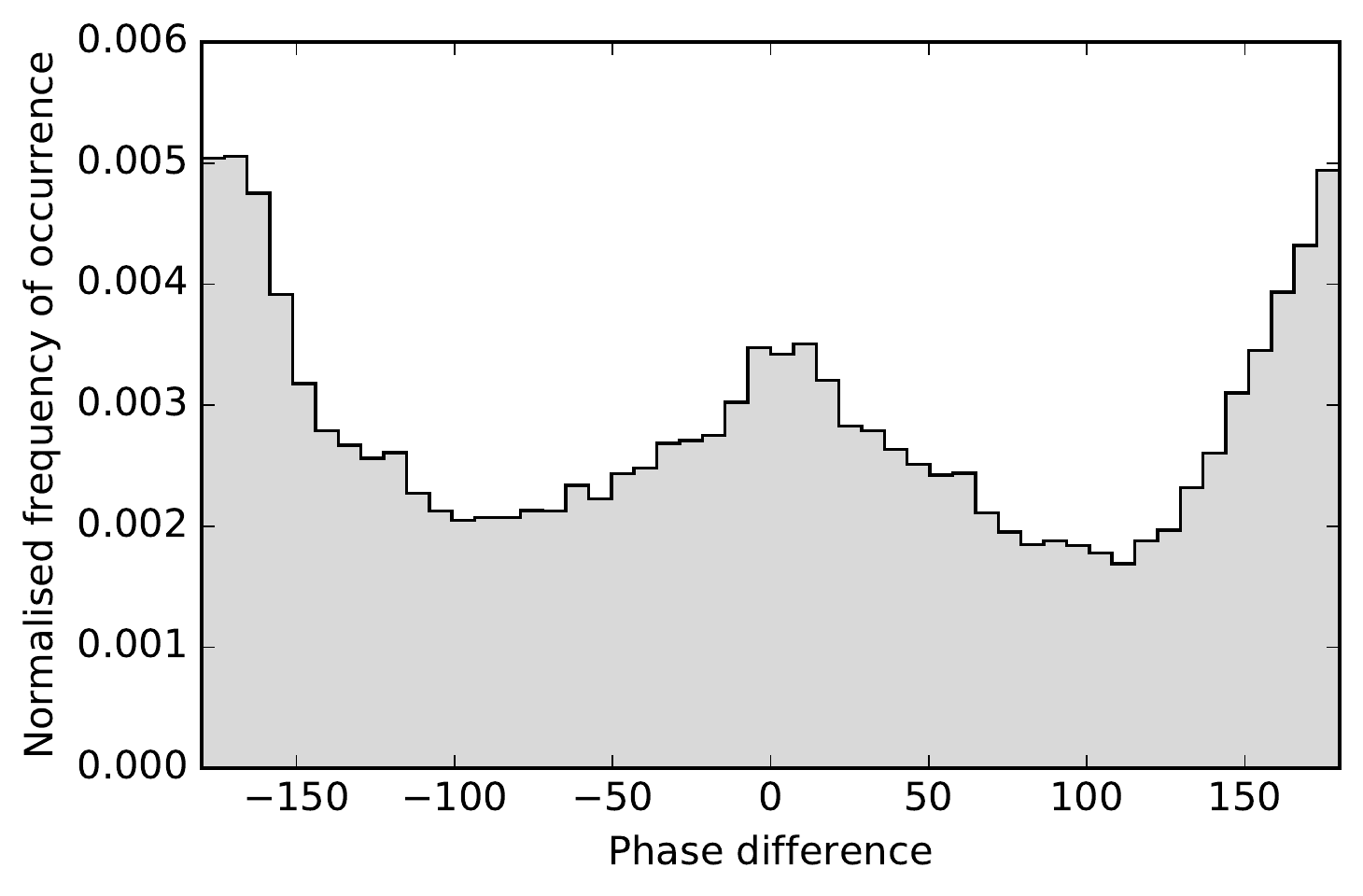}
   \caption{Distribution of phase differences between width and intensity oscillations in the SCFs at a given cut across each fibril.}
   \label{phasehist}
\end{figure}

\begin{figure}[ht]
   \centering
   \includegraphics[width=9cm]{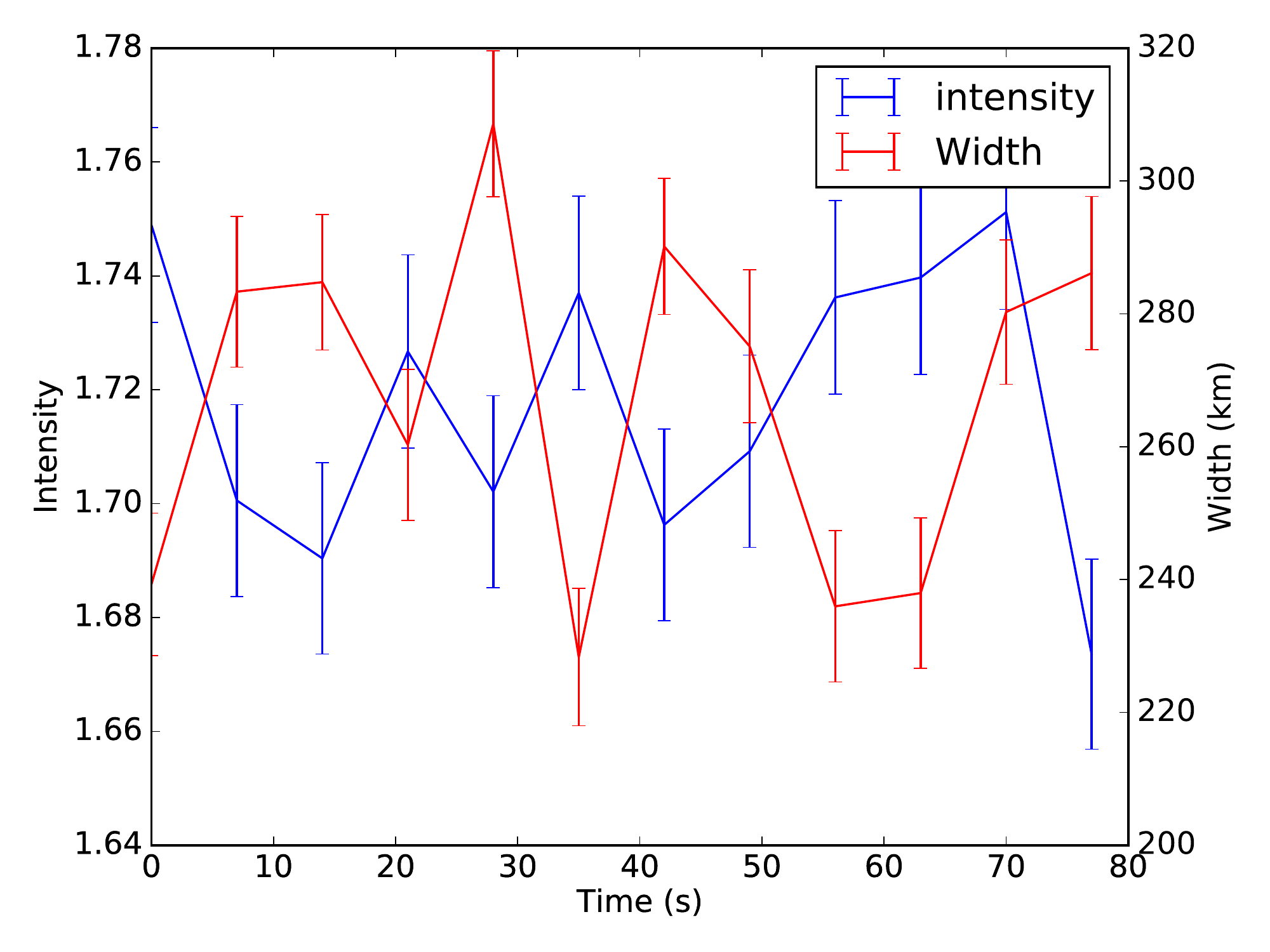}
   \caption{An example of a clear anti-correlation between oscillations in maximum intensity (blue line) and width (red line) of the sample SCF in \fig{mesh_mp_w_ex1}, at the location marked by the white arrow in the lowest panel of that figure.
   The error bars represent the standard deviations of the photon counts for the intensity, and the uncertainties of the Gaussian fitting to the cross-section of the fibril for the width.}
   \label{int_wid}
\end{figure}

We computed the wavelet cross-power spectra between width and intensity oscillations within a given cut across a SCF and determined the phase lag between the oscillations in these quantities. A wide range of phase lags was obtained. \fig{phasehist} illustrates the distribution of values of this quantity. The distribution has two clear peaks, a weaker one at 0$^\circ$ (in phase), and strong peak at $\Mypm{} 180^\circ$ (anti phase).
As an example for such an anti-phase oscillation we present in \fig{int_wid} the temporal evolution of fibril width and intensity of the sample SCF at the position indicated by a white arrow in \fig{mesh_mp_w_ex1}.

\section{Discussion and conclusions}
\label{sec:conclusions}
We have provided observational evidence for oscillations of the width of fibrils and of intensity along slender \caiih{} fibrils (SCFs) in the lower chromosphere.

The high-resolution, seeing-free images under study were recorded with \sunrise/\sufi{} and revealed that such oscillations are almost ubiquitous all over the field of view, which covered part of NOAA AR 11768.
Fluctuations in length of some of the SCFs were also observed, although it is not clear how independent this parameter is from the intensity.

The oscillatory behavior of both parameters (i.e., fibril width and intensity) was identified in wavelet power spectra determined at a series of locations along the backbone of each detected SCF. The wavelet transform was employed to analyze the fluctuations, from which median periods of $32\pm17$\,s and $36\pm25$\,s were obtained for the width and intensity oscillations, respectively, with the uncertainty intervals representing the standard deviations of each distribution. Cross power spectra between the perturbations at different locations along a given fibril revealed phase speeds of ${11}^{+49}_{-11}$\,\kms{} and ${15}^{+34}_{-15}$\,\kms{} for the width and intensity fluctuations, respectively. Again, uncertainty intervals reflect standard deviations.
Given that these waves display brightness and width signatures, they have to be compressible, ruling out Alfv\'en waves. Simultaneous periodic fluctuations of intensity and width in elongated structures are manifested by either slow-mode waves or fast sausage-mode waves \citep{Van-Doorsselaere2011,Su2012}. In addition to observations of the latter wave mode in the upper solar atmosphere \cite[i.e., in the upper chromosphere and in the corona,][]{Inglis2009}, observations of sausage oscillations have also been reported at lower atmospheric heights, in structures such as pores \citep{Dorotovic2008,Morton2011}.

To distinguish which wave modes are present in the SCFs we need to compare the phase speeds of the oscillations with the expected plasma Alfv\'en and sound speeds. The values for the Alfv\'{e}n and sound speeds were computed using the NC5 flux-tube from \citep{1993A&A...273..293B} embedded in the VAL-A atmosphere \citep{1981ApJS...45..635V}.
For a majority of the detected waves these velocities are larger than the sound speed at the low chromospheric heights sampled by the \sufi{} 1.1\,\AA{} \caiih{} filter, which lies around 7\,\kms{}. The measured phase speeds are comparable to the local Alfv\'{e}n speed for this height region, with typical values in the range of 7--25\,\kms{}.

Interestingly, 20\% of the SCFs show phase differences between $-30^\circ$ and $+30^\circ$, indicative of in-phase oscillations.
A strong peak was found at $\Mypm180^\circ$, with about 50\% of all oscillations having a phase difference within 30$^\circ$ of $180^\circ$ (anti phase oscillations). Phase differences of around $\pm 90^\circ$ are relatively uncommon.

A phase difference in the range of $\left|150^\circ-180^\circ\right|$, as displayed by the example shown in \fig{int_wid}, is consistent with the signature of sausage-mode oscillations in the SCFs under the assumption of an optically thin plasma. The validity of this assumption is confirmed by the fact that the observed intensity increases at the intersection points of crossing SCFs, suggesting that we can partly see through individual fibrils. 
The contraction of the fibril caused by the sausage-mode oscillation leads to a narrow fibril with a higher density. In an optically thin regime, a higher density implies an increased intensity. The subsequent expansion phase of the oscillation leads to an increase of the fibril's width at lower intensity.
For such a plasma, the intensity follows the electron density. However, a detailed model of the brightening of these structures is required to determine the behaviour of the plasma under conditions typical of the lower chromosphere.

In about 25\% of our SCFs we did not find a clear correlation between the fluctuations in the two parameters. The intensity oscillations could also be caused by slow mode waves, which are expected to be present inside strong-field magnetic features, such as flux tubes (although it is unclear to what extent the SCFs can be described as flux tubes). However, the median phase speeds obtained in our analysis are too high for slow mode waves. Only for a few SCFs do we obtain low phase speeds that may well be compatible with the slow mode.

To our knowledge, our observations of sausage mode oscillations in the SCFs are the first direct evidence of this wave mode in the lower solar chromosphere. \citet{Morton2012} inspected oscillations of width and intensity in H$\alpha$ elongated fibrils and short mottles (in the upper chromosphere). They found a phase speed of $67\pm15$\,\kms{} for their MHD fast sausage waves, which is much larger than those we found in the SCFs. Like us, they also found a phase difference of 180$^\circ$ between their detected intensity and width perturbations. \citet{Dorotovic2008} and \citet{Morton2012} showed that the energy these sausage waves carry is sufficient to contribute (around 10\%) to the heating of the chromosphere and/or the corona. \citet{Jess2012b} claim to see a fluctuation of the width of what they call a chromospheric spicule (observed on the disk as an H$\alpha$ dark fibril). They interpret these fluctuations as sausage modes in the chromosphere. 

This work points to a number of follow-up investigations to  advance our knowledge and understanding of the detected oscillations and waves. Firstly, measurements that include velocities, would help to distinguish better between different possible wave modes. Another important step is to compute MHD wave modes in simple models of fibrils, possibly described as flux tubes embedded in a magnetized gas. Such a study should not only lead to new insights into the physics of these oscillations, but would also reveal the expected behaviour of different physical parameters, thus providing guidance for future observations and their interpretation. Finally, an investigation of the  physical processes that drive this oscillatory behavior of the SCFs would also be very useful.

\begin{acknowledgements}
The German contribution to \sunrise{} and its reflight was funded by the Max Planck Foundation, the Strategic Innovations Fund of the President of the Max Planck Society (MPG), DLR, and private donations by supporting members of the Max Planck Society, which is gratefully acknowledged. The Spanish contribution was funded by the Ministerio de Econom\'{\i}a y Competitividad under Projects ESP2013-47349-C6 and ESP2014-56169-C6, partially using European FEDER funds. The HAO contribution was partly funded through NASA grant number NNX13AE95G. This work was partly supported by the BK21 plus program through
the National Research Foundation (NRF) funded by the Ministry of Education of Korea.
SJ receives support from the Research Council of Norway.
\end{acknowledgements}

\bibliographystyle{aa}

\end{document}